\documentclass[12pt,singlecolumn,twoside]{osajnl}
 \setboolean{shortarticle}{true}
\journal{ol}

\usepackage{multirow}
\usepackage{xcolor}
\usepackage[textwidth=1.0 cm]{todonotes}

\title{Tornado waves}

\author[1,2]{Apostolos~Brimis}
\author[1,2]{Konstantinos~G. Makris}
\author[1,3*]{Dimitris~G.~Papazoglou}

\affil[1]{Institute of Electronic Structure and Laser (IESL), Foundation for Research and Technology - Hellas, P.O. Box 1527, GR-71110 Heraklion, Greece}
\affil[2]{ICTP, Department of Physics, University of Crete, Heraklion, Greece}
\affil[3]{Materials Science and Technology Department, University of Crete, GR 70013, Heraklion, Greece}
\affil[*]{Corresponding author: dpapa@materials.uoc.gr}

\begin{abstract}
We show that light spiraling like a tornado can be generated by superimposing abruptly auto-focusing ring-Airy beams that carry orbital angular momentum of opposite handedness. {With different parabolic propagation trajectories, the superimposing ring-Airy beams are tailored to abruptly auto-focus at overlapping focal regions}. This results to a complex wave with intense lobes that twist and shrink in an accelerating fashion along propagation. 
{By achieving angular acceleration values that exceed  $295 ~\text{rad} / \text{mm}^2$, these Tornado waves can find numerous applications in laser trapping, direct laser writing and high harmonic generation}.
\end{abstract}

\setboolean{displaycopyright}{false}
\begin{document}

\maketitle

{Shaping an optical wave packet to realize a strong focus along propagation in linear or nonlinear media is a topic of wide interest in optics. The cylindrically symmetric accelerating beams, often referred as ring-Airy beams or circular Airy beams (CAB), that were recently introduced \cite{Efremidis2010a,Papazoglou2011,Efremidis2019,Penciu2016} represent such kind of tailored waves}. These shaped wave packets abruptly auto-focus and propagate in curved trajectories, while at high intensities they reshape into nonlinear intense light-bullets with extremely well defined focal position \cite{Panagiotopoulos2013}. In the same context, by imprinting a helical phase, shaping can induce topological charge to the wave packet. Such shaped wave packets, often referred as optical vortices, carry orbital angular momentum (OAM) and exhibit a rotating phase structure as they propagate \cite{Gauthier2017,Yao2011}. Furthermore, the phase singularity in the  vortex center leads to a donut-shaped intensity profile \cite{Gauthier2017,Davis2012,Yao2011} with various applications in microscopy and optical trapping among others \cite{Yao2011}. 
\begin{figure}[!t]
  \centering
  \includegraphics[width=0.6\textwidth]{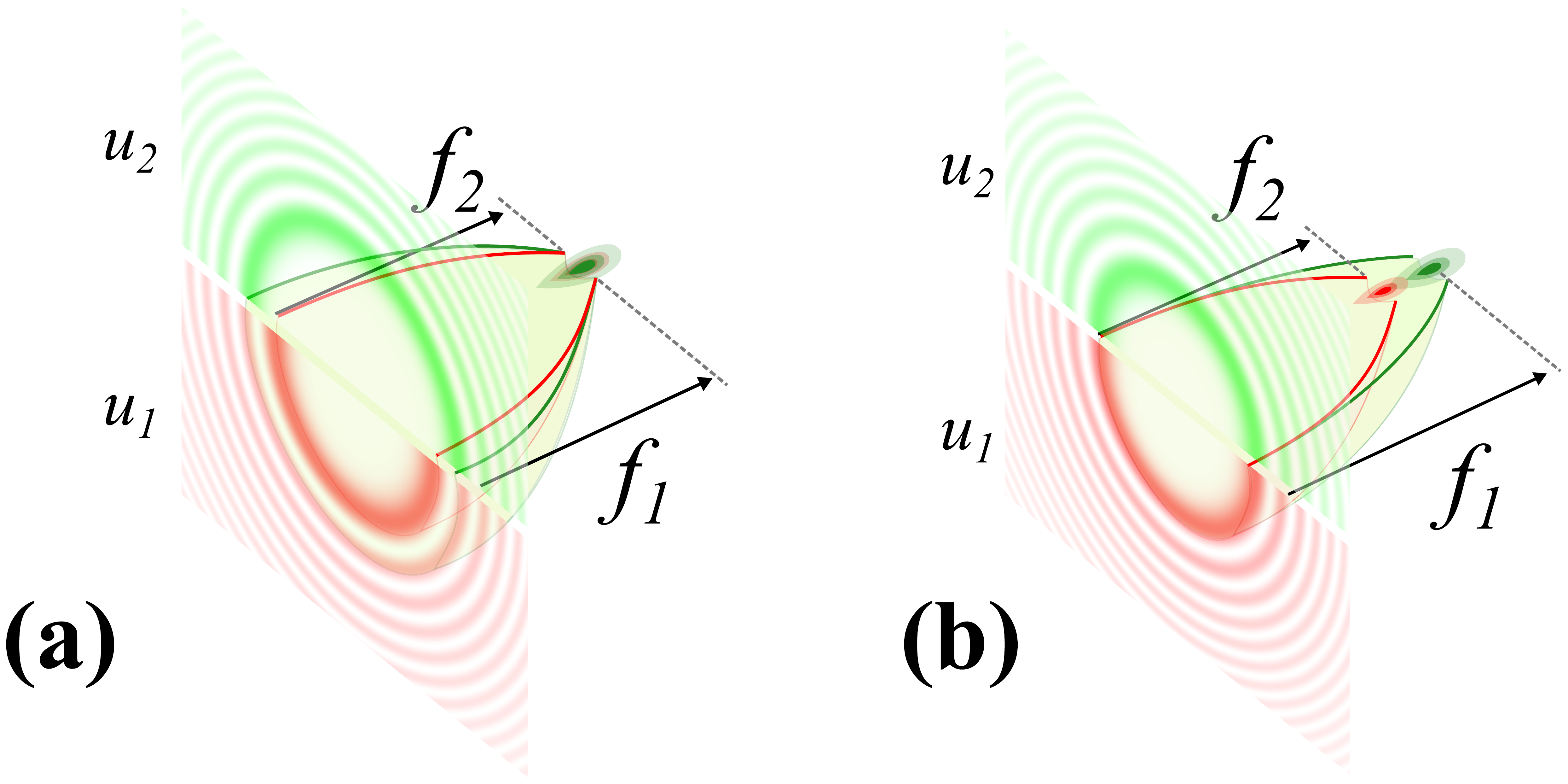}
\caption{Graphical depiction of the interference between two ring-Airy beams for different beam parameters. (a) the foci coincide ($r_2>r_1, w_2<w_1$), (b) the foci partially overlap  ($r_2=r_1, w_1<w_2$). The intensity profile of $u_1, u_2$ at the initial plane $(z=0)$ is shown in red and green respectively.}
\label{fig:TornadoTailoring}
\end{figure}
On the other hand, the interference of structured light that caries OAM has recently attracted a lot of attention with a variety of beam configurations being studied. For example, in the case of two superimposing Bessel beams~\cite{Vasilyeu2009,McGloin2003,Rop2012} the intensity pattern rotates at a constant rate, forming a helix as it propagates. The first realization of angularly accelerating light was reported by Schulze et al. \cite{Schulze2015} by superimposing two pairs of complex beams. Each pair was consisting of two Bessel beams carrying OAM of opposite helicity,  while  the Bessel conical angle varied between the pairs. This twisting light rotates in a tailored accelerating, or decelerating, fashion forming a helix of variable pitch as it propagates \cite{Schulze2015}. In a similar fashion,  by superimposing two Laguerre-Gaussian beams with slightly different Rayleigh lengths, opposite helicities and same radial dependence,  a radially dependent angular accelerating light is generated \cite{Webster2017}. This angular acceleration is localized only around the focus where it is notable. 
\begin{figure}[t]
  \centering
  \includegraphics[width=0.6\textwidth]{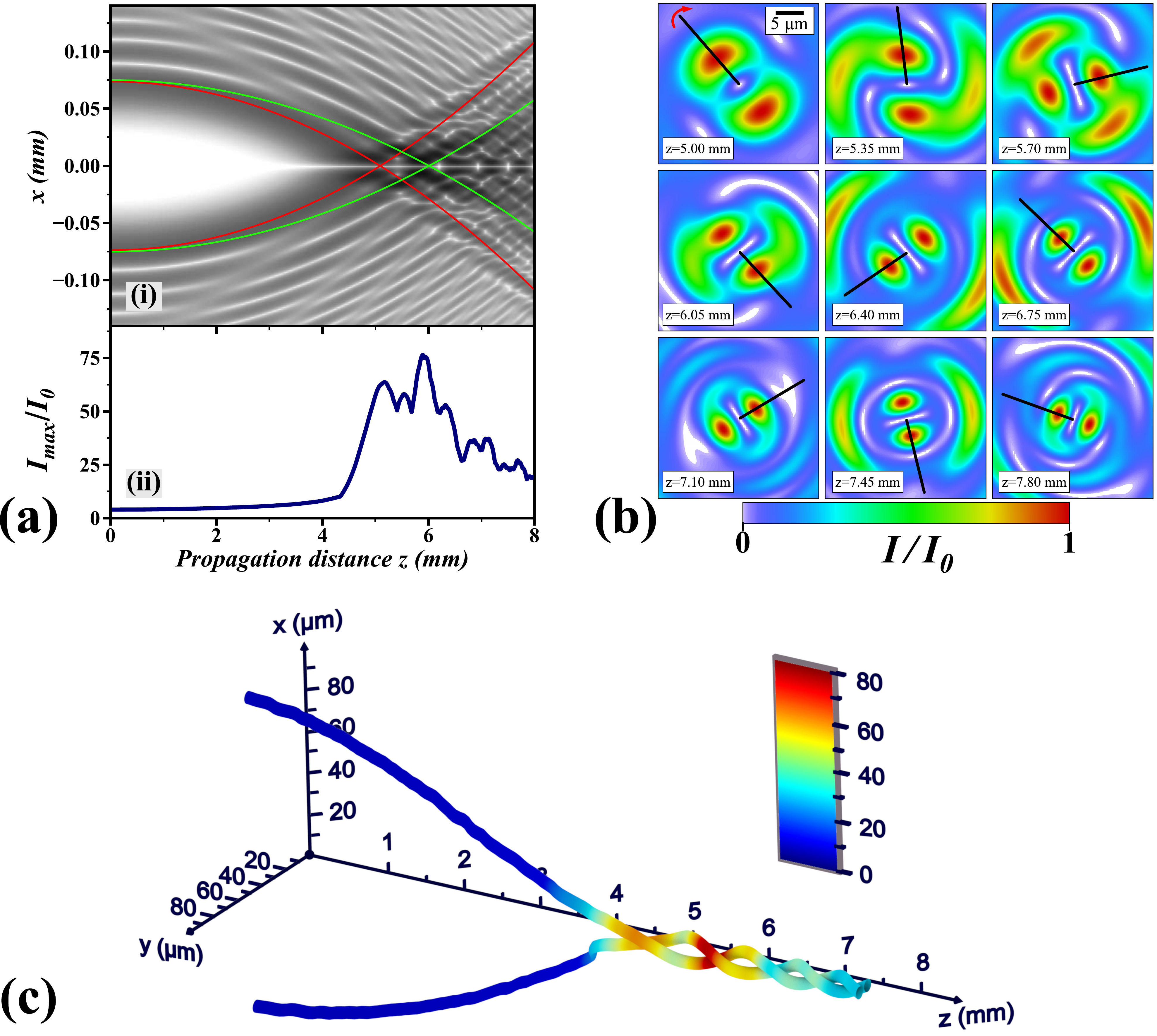}
\caption{ Superposition of two co-propagating interfering vortex ring-Airy beams; (a) (i) cross sectional intensity profile $I(x,z)$  (ii) Intensity contrast $I_{c}(z)=I_{max}(z)/I_0$  as function of the propagation distance $z$, cross sectional intensity profile $I(x,z)$. (The green and red curves depict the parabolic trajectory of each beam) (b) transverse $I(x,y)$ intensity profiles along the beam propagation ($\Delta z=350\mu\text{m}$) (the black lines are guides to the eye)  (c) Visualization of the spiraling trajectory of the high intensity lobes.}
\label{fig:Interfering_RA_Beams}
\end{figure} 

In this work we show that it is possible to generate light that twists and accelerates both over the radial and the angular direction, that we define as Tornado waves (ToWs). These waves exhibit intensity maxima that, like a tornado, outline a spiral of decreasing radius and pitch as they propagate. We achieve to generate ToWs by superimposing two abruptly auto-focusing ring-Airy beams that are tuned to overlap their focal regions while they carry OAM of opposite helicity. 

The superposition of a two ring-Airy beams carrying OAM of opposite helicity can be described by
\begin{gather}
\label{eq:tornadodef}   
 u\left( {r,\varphi } \right)=u_1\left( {r,\varphi }\right)+u_2\left( {r,\varphi }\right),\\ \notag
u_i\left( {r,\varphi } \right)=u^o_i Ai({\rho _i}){e^{a{\rho _i}}}{e^{i l_i\varphi }}
\end{gather}
where $Ai$ is the Airy function, $r, \varphi$ are respectively the radial and azimuthal coordinates, ${\rho_i} = \left( {{r_i} - r} \right)/{w_i}$, $r_i$ are the radius and $w_i$ the width parameters of the primary ring \cite{Papazoglou2011,Panagiotopoulos2013}, $u^o_i$ is the beam amplitude, $l_i$ is the topological charge, and $a$ is an apodization factor. The propagation of such beams is described by the paraxial wave equation    \cite{Efremidis2010a,Papazoglou2011}:
\begin{gather}
\nabla_{\bot}^2 u + 2 i k \frac{\partial {u}}{\partial z}  = 0
\label{eq:paraxial}    
\end{gather}
where $\nabla_{\bot}^2$ denotes the transverse part of the Laplacian, $k$ the free space wavenumber, $z$ the propagation distance and $u$ the electric field envelope. Eventhough no analytical solutions of Eq. \ref{eq:paraxial} 
for the propagation of ring-Airy beams are known to exist, we can still predict the position of their abrupt autofocus. In particular, based on the analytical solution of the one-dimenisonal Airy beam \cite{Siviloglou2007,Siviloglou2007a}, each of the interfering ring-Airy beams will abruptly autofocus at ${f_i} = 4{z_i}\sqrt {{r_i}/{w_i} + 1}$, where ${z_i \equiv k{w_i^2}/2}$  ~\cite{Papazoglou2011,Mansour2018,Panagiotopoulos2013}. As shown schematically in Fig.~\ref{fig:TornadoTailoring} we can control the overlap region of the two beams by tuning the values of $r_i,~w_i$ and thus tailor the ToW behaviour.

\begin{table}[!b]
\centering
\caption{{Tornado wave parameters}}
\label{Tab:Overview}
\begin{tabular}{ccccccc}
\hline
\multirow{2}{*}{ToW} & $r_1$           & $w_1$           & $r_2$           & $w_2$           & $a$  & $\lambda$ \\
                     & $(\mu\text{m})$ & $(\mu\text{m})$ & $(\mu\text{m})$ & $(\mu\text{m})$ &      & $(nm)$    \\ \hline
A                    & 62.5            & 12.5            & 62.5            & 11.25           & 0.03 & 800       \\
B                    & 49.05           & 11.9            & 109.52          & 9.52            & 0.03 & 800       \\ \hline
\end{tabular}
\end{table}
In order to study the properties of ToWs we performed numerical simulations of  Eq.~\ref{eq:paraxial}
for light fields linearly polarized  along the $\textbf{x}$ direction. As an example of a Tornado wave we choose the scheme shown in Fig.~\ref{fig:TornadoTailoring}(b) where the foci of $u_1, u_2$ partially overlap {(using the parameters of set A in Table~\ref{Tab:Overview})} while their  topological charges are respectively $l_1=-l_2=1$.  As shown in the cross-sectional plot of Fig.~\ref{fig:Interfering_RA_Beams}(a) the superposition of the two ring-Airy beams preserves the characteristic parabolic trajectory and the abrupt auto-focus \cite{Efremidis2010a,Papazoglou2011}. In this scheme, as they propagate, the two interfering beams follow slightly different trajectories while their primary rings are partially overlapping. This generates, as shown in Fig.~\ref{fig:Interfering_RA_Beams}(b), a rotating intensity pattern in the transverse plane along the propagation direction. The trajectory of the high intensity lobes is visualized in Fig.~\ref{fig:Interfering_RA_Beams}(c). Resembling  a tornado, the trajectories form a spiral of decreasing radius and pitch as the beam propagates. This is a clear demonstration of a Tornado wave where light twists and accelerates both over the radial and the angular direction. The parabolic trajectory up to the focus is related to radial acceleration, while the decreasing pitch ~\cite{Schulze2015,Webster2017} is related to the angular acceleration.
\begin{figure}[!t]
  \centering
  \includegraphics[width=0.6\textwidth]{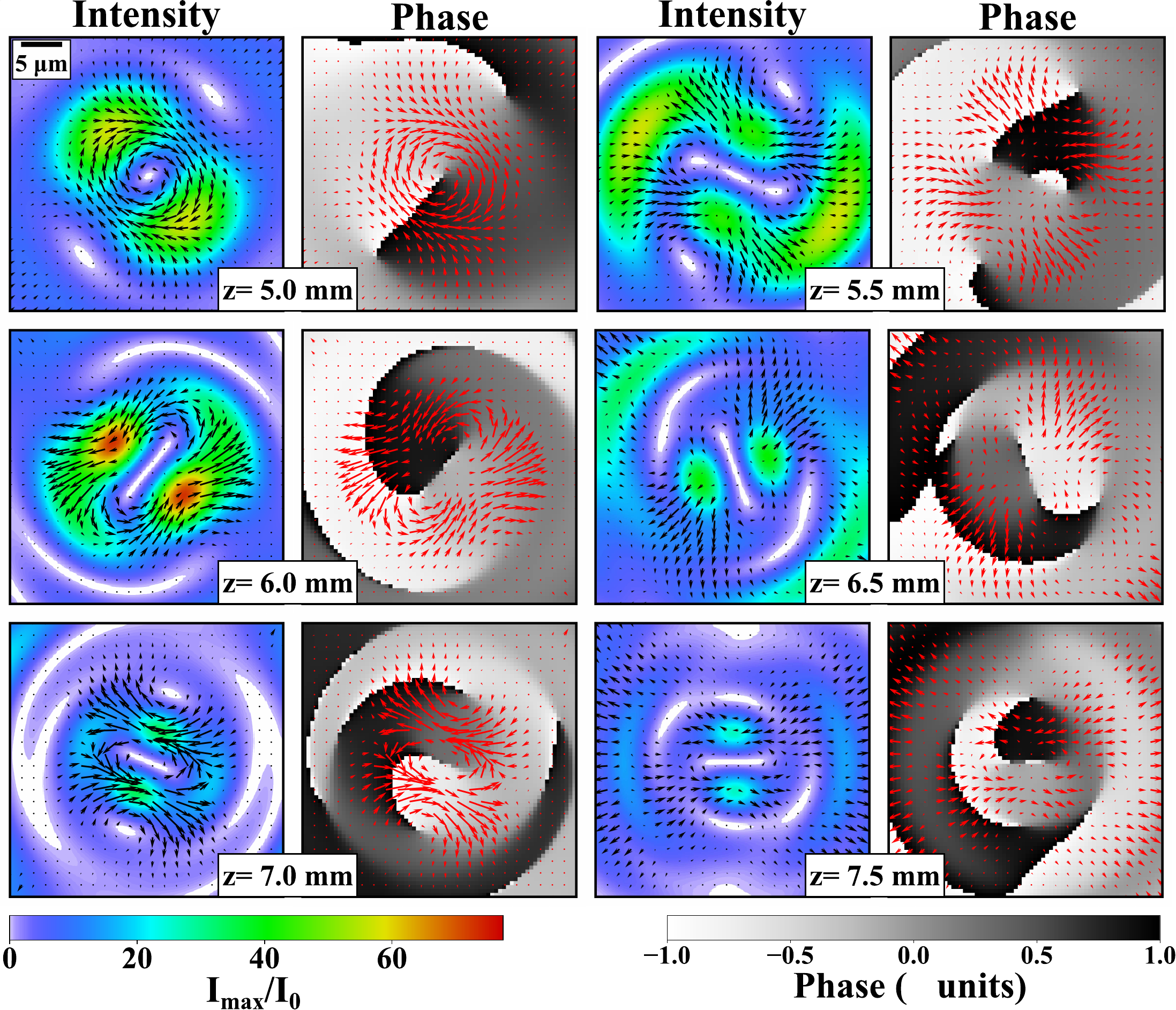}

\caption{Cross sectional images of intensity (false colors) and phase (inverted gray, wrapped in $2\pi$) of a ToW at different z planes ($\Delta z=500\mu\text{m}$). Arrows (black and red respectively) represent the transverse component of the Poynting vector $\mathbf{S}_{\bot}$.}
\label{fig:Int_Phase}
\end{figure}
\begin{figure}[!b]
  \centering
  \includegraphics[width=0.6\textwidth]{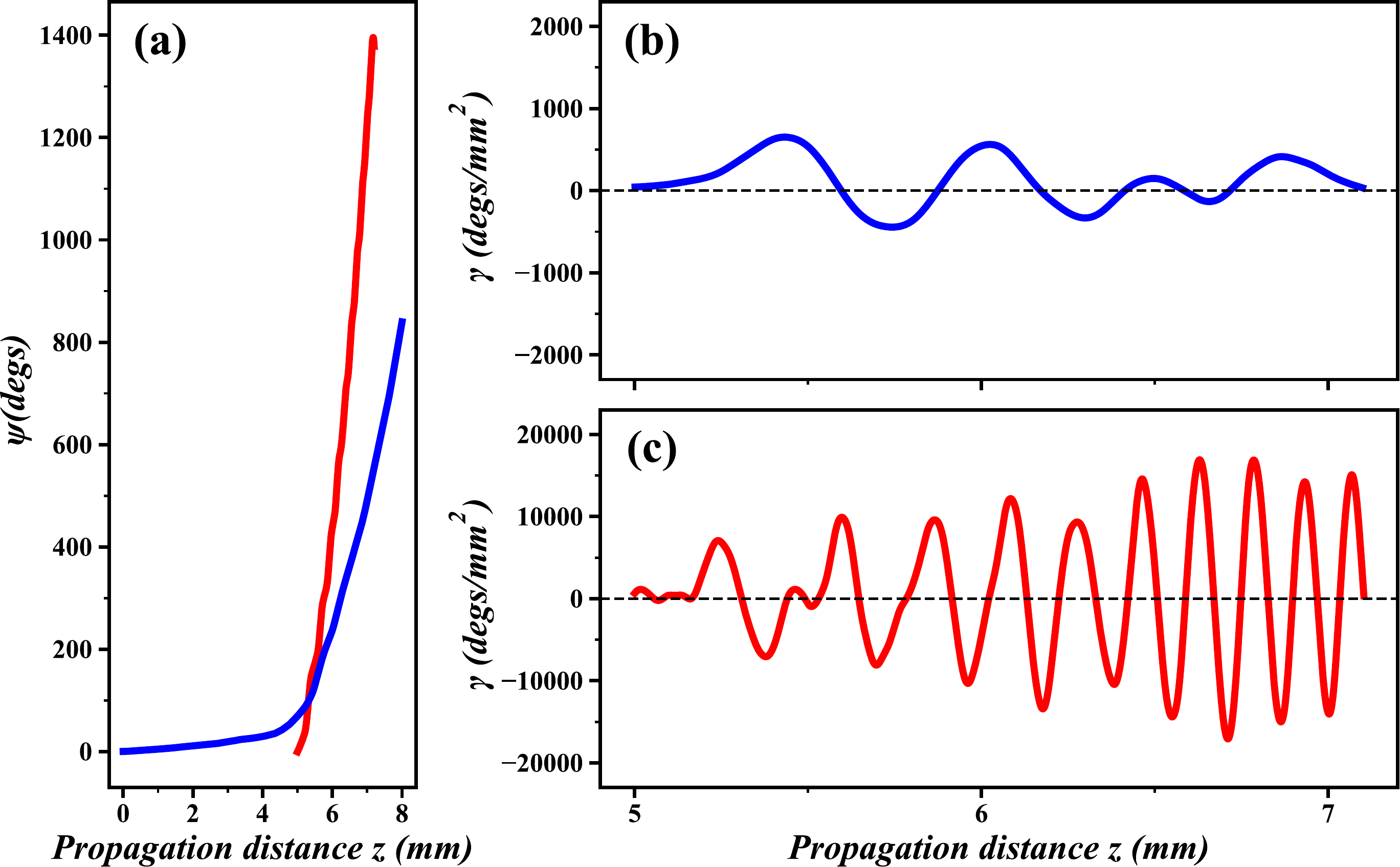}
\caption{(a) angular position $\psi (z)$ for partial (blue dotted curve), and complete (red solid curve) foci overlap (see Fig. \ref{fig:TornadoTailoring}). (b),(c) angular acceleration $\gamma$ of the high intensity lobes as a function of the propagation distance for the partial (b) and complete foci overlap (c).}
\label{fig:Theta_Acc}
\end{figure}
{The power flow of a propagating optical wave is described\cite{Born1999} by the Poynting vector $\bf{S}$. The power flow normal to the propagation axis can be then obtained by the transverse component of the Poynting vector ${\bf{S_{\bot}}}$. At the paraxial limit, for linear polarization,  this is expressed as ${\bf{S_{\bot}}}  =   \frac{i}{2 \omega \mu_0} (u\nabla_{\bot} u^{*}-u^{*}\nabla_{\bot} u)$, where $\omega$ is the angular frequency of light and $\mu_0$ the permeability of free space}. The intensity and phase of the ToW at different propagation distances is presented in Fig.~\ref{fig:Int_Phase} along with the power flow (Poynting vector) \cite{Allen} in the transverse $xy$ plane. Clearly, although the total OAM is zero, vortices are generated and annihilated along the propagation direction. Vortices are better visualized at the cross sections of the ToW phase, where for example three vortices, with one in the center, can be clearly seen at $z=5\text{mm}$, while two vortices are seen at $z=7\text{mm}$. 
\begin{figure}[!b]
  \centering
  \includegraphics[width=0.6\textwidth]{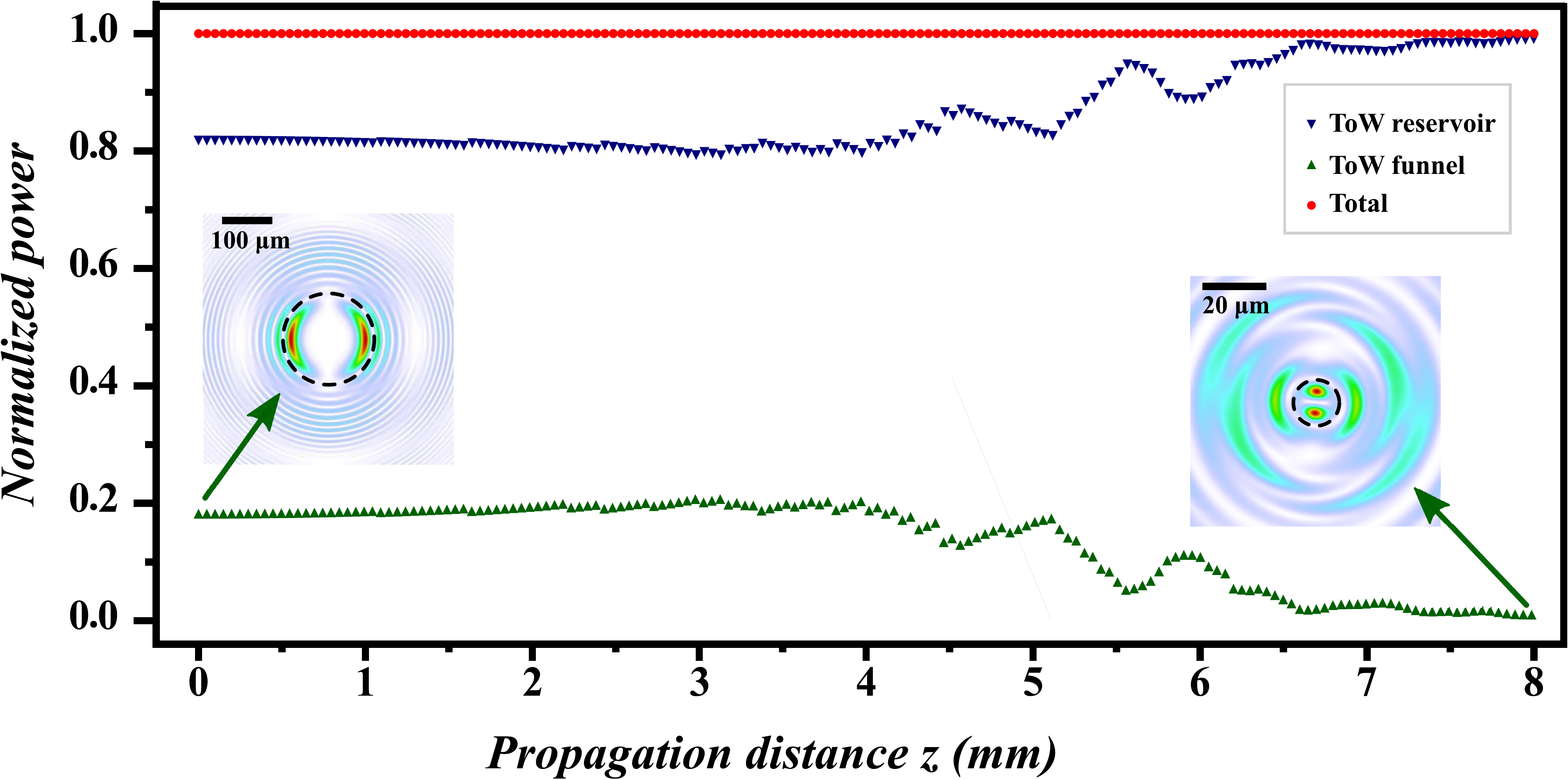}
\caption{Normalized power as a function of the propagation distance of the beam. Insets show the evolution of the ToW funnel}
\label{fig:EnergyConsrv}
\end{figure}
Likewise, we have traced the angular position $\psi(z)$ of the high intensity lobes as a function of the propagation distance by analyzing the cross sectional $I(x,y)$ profiles. From these values we have estimated the angular velocity $\dot \psi(z)$ and the angular acceleration $\gamma \equiv \ddot \psi(z)$. {We have studied two cases of ToWs, one with partial foci overlap, and one with complete foci overlap using the parameters of sets A and B in in Table~\ref{Tab:Overview} respectively}. As shown in Fig.~\ref{fig:Theta_Acc}(a) the high intensity lobes exhibit at least two complete rotations for a propagation distance of $~8~\text{mm}$. In the case of complete foci overlap  $\psi(z)$ values start from the focus since, as it can be seen in Fig. \ref{fig:TornadoTailoring}(a), the $u_1,u_2$ ring-Airy beams do not overlap before that point and thus there are no intensity peaks.  The non-linear shape of the curves indicates that these rotations take place in an accelerating fashion. This is confirmed in Fig.~\ref{fig:Theta_Acc}(b),(c) where the angular acceleration is shown to reach values of $650 \deg/\text{mm}^2$ and $1.69 ~ 10^4 \deg/\text{mm}^2$ for the case of partial and complete foci overlap respectively. Clearly when the two ring-Airy beams carrying OAM of opposite helicity have overlapping foci the angular acceleration is optimized reaching values that are by more than 24 times higher compared to the partially overlapping case. We should note here that, as we are going to discuss in detail later, ToWs exhibit angular acceleration values that  are orders of magnitude higher than those reported so far  \cite{Schulze2015,Webster2017}. Another important aspect is the amount of beam power that is carried by the twisting high intensity lobes of the ToW. The normalized power carried by the two twisting lobes (see Fig.~\ref{fig:Interfering_RA_Beams}) which we refer to as ToW funnel region and that of the rest of the beam, which we refer to as reservoir, is presented in Fig.~\ref{fig:EnergyConsrv}. Although their peak intensity is many times greater than the initial~($20~I_0$), the twisting lobes carry a decaying portion of the total beam power as they propagate. Furthermore, the oscillatory behaviour observed after the focus in $~4.5\text{mm}$ is due to the rotational acceleration of the twisting lobes. Such an acceleration is related to the energy exchange between the ToW funnel and the reservoir regions.  
\begin{figure}[!t]
  \centering
  \includegraphics[width=0.6\textwidth]{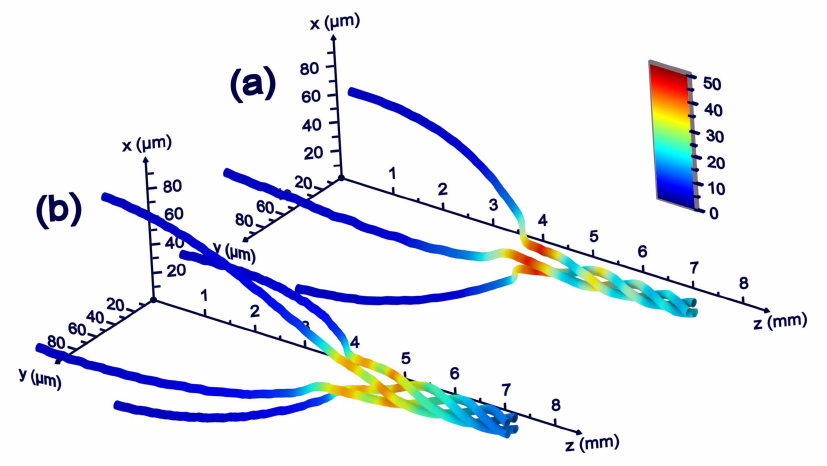}
\caption{Visualization of the trajectory of high intensity lobes for Tornado waves,  generated by the superposition of ring-airy beams {(set A, Table~\ref{Tab:Overview})} carrying OAM of topological charge (a) $l_1=2,~l_2=-1$ (b) $l_1=2,~l_2=-2$ }
\label{fig:VariousOAM}
\end{figure}
\begin{figure}[!b]
  \centering
  \includegraphics[width=0.6\textwidth]{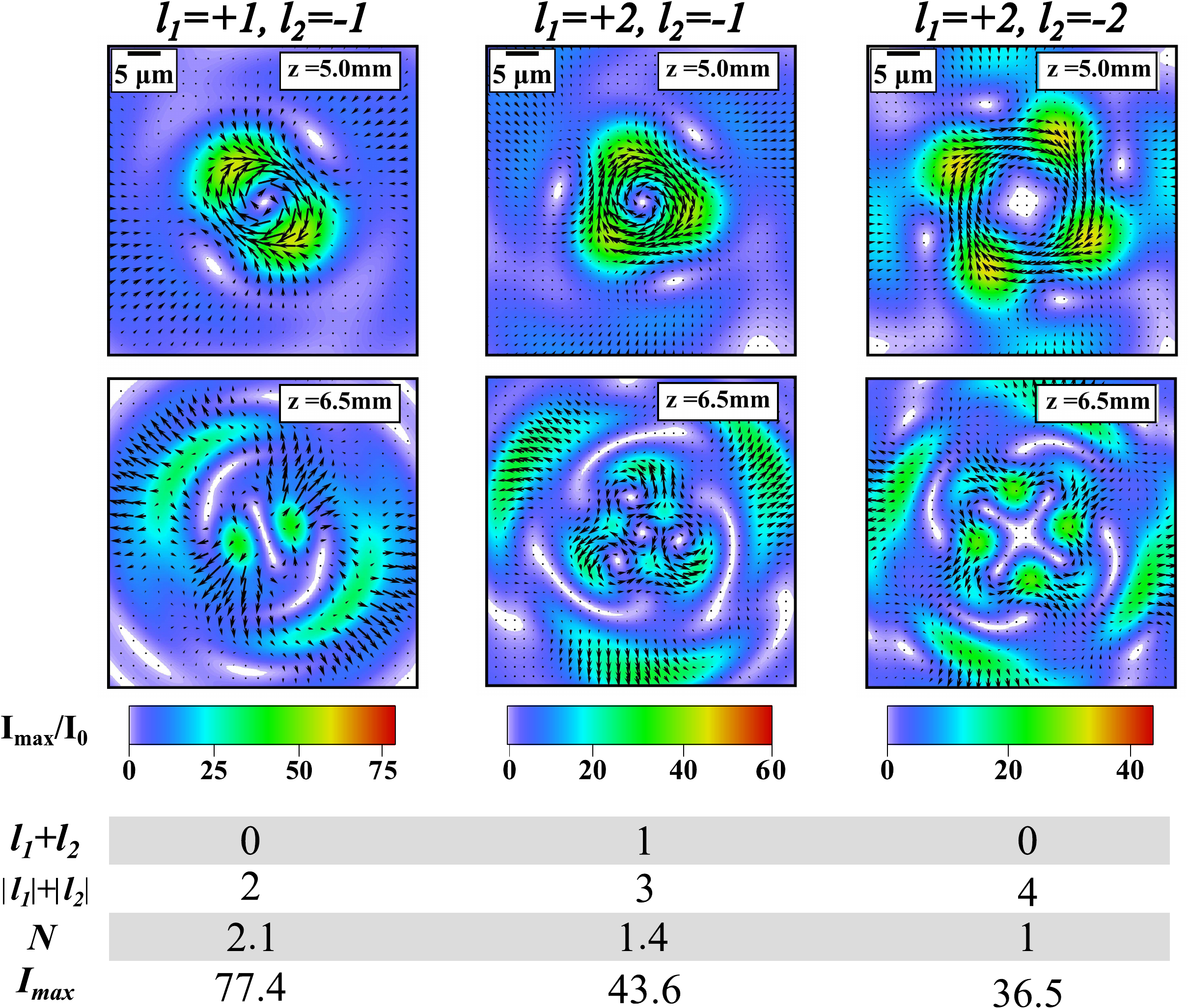}
\caption{ Intensity $I(x,y)$ cross-sectional images for various combinations of topological charges. ($l_1+l_2$ refers to the total OAM per photon at $\hslash$ units,  $|l_1|+|l_2|$ refers to the total number of intensity peaks, and $N, I_{max}$ refer to the number of turns and the intensity contrast in the focal region). Detailed animations of the propagation for $(l_1=+1,l_2=-1)$, $(l_1=+2,l_2=-1)$ and $(l_1=+2,l_2=-2)$  are respectively shown in  Visualization 1,  Visualization 2, and Visualization 3.}
\label{fig:Various_lobes}
\end{figure}

Using the same approach we can generate ToWs by superimposing ring-Airy beams carrying OAM with various combinations of topological charges {for the case of partial foci overlap (parameters of set A in Table \ref{Tab:Overview})}. In Fig.~\ref{fig:VariousOAM} the trajectory of the high intensity lobes for such combinations is shown. An increasing number, equal to $\left| {{l_1}} \right| + \left| {{l_2}} \right|$, of high intensity lobes twist shrinking in radius and pitch. Both in the case where the total OAM is not zero (see Fig.~\ref{fig:VariousOAM}(a) where $l_1=2,~l_2=-1$) and in the case of a zero total OAM  (see Fig.~\ref{fig:VariousOAM}(b) where $l_1=2,~l_2=-2$) the tordado-like shape is profound. This confirms that ToWs are generated from the superposition of any combination of accelerating abrupt auto-focusing waves carrying OAM of opposite handedness, giving another degree of freedom for tailoring their properties. Finally, we have summarized in Fig.~\ref{fig:Various_lobes} the behaviour of the tornado wave for different combinations of topological charge and for different propagation distances. Likewise, animations of the propagation for $(l_1=+1,l_2=-1)$, $(l_1=+2,l_2=-1)$ and $(l_1=+2,l_2=-2)$  are respectively shown in  Visualization 1,  Visualization 2, and Visualization 3. {In all cases the OAM of the whole beam ($\hslash$ per photon) is conserved. On the other hand, as we observe vortices are annihilated/generated along propagation thus the OAM is not  conserved locally.}

As the values of $l_1$, $l_2$ increase, a more complex intensity pattern is formed. For example, in the case of $l_1=2$, $l_2=-1$ (second column of Fig.~\ref{fig:Various_lobes}), besides the $|l_1|+|l_2|=3$ primary lobes we observe a complex structure of secondary lobes. Furthermore, as the number of lobes increases, the maximum intensity and the the total angular rotation $\psi_{max}$ at a specific propagation distance decreases.

\begin{figure}[!b]
  \centering
  \includegraphics[width=0.6\textwidth]{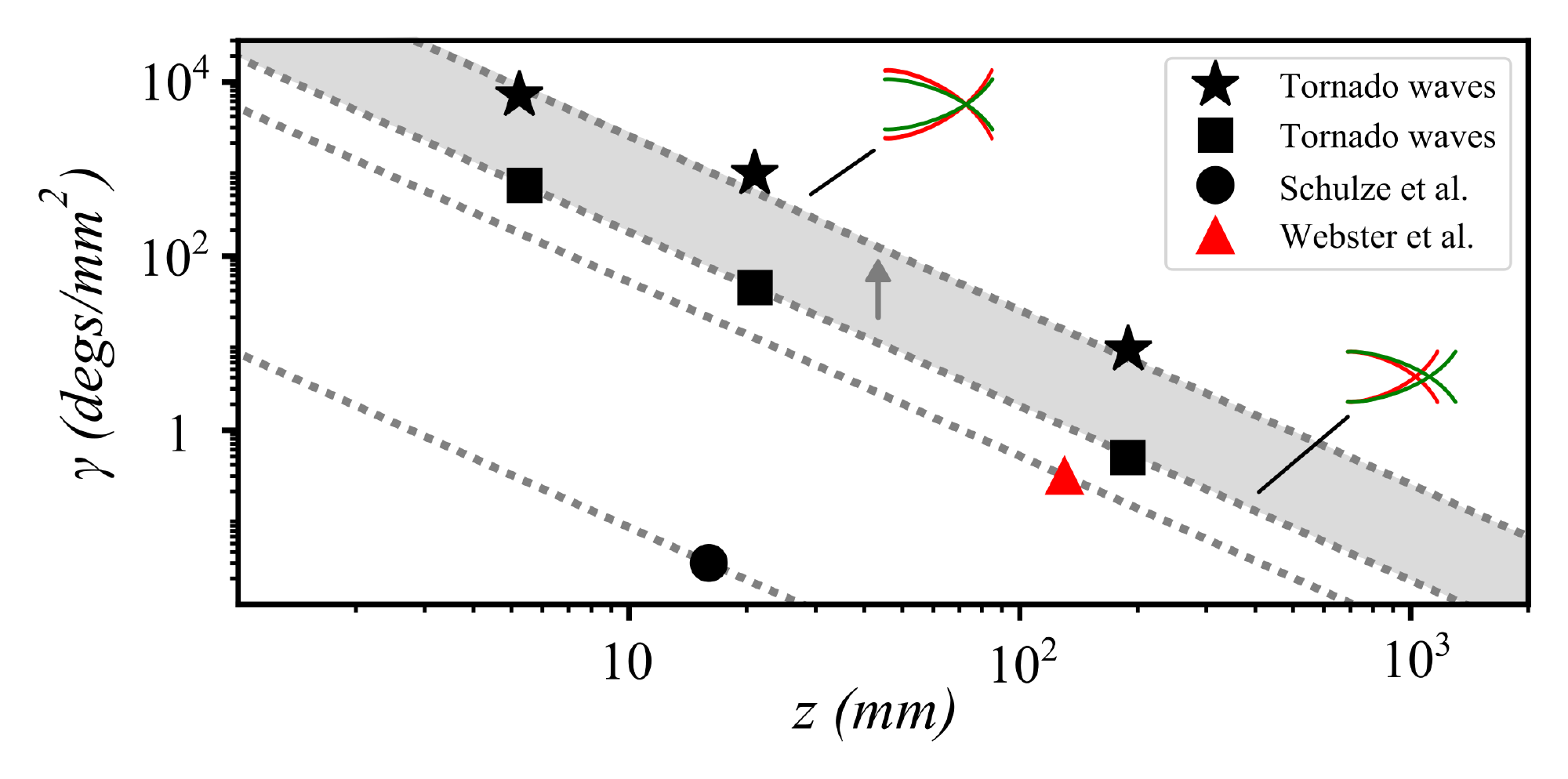}
\caption{Angular acceleration values for ToWs at various focusing distances. The gray zone denotes the transition from partial to complete overlap of the foci as indicated by schematics.  Results from Webster et al. \cite{Webster2017} and Schulze et al. \cite{Schulze2015} are also shown for comparison. The dotted lines denote an $1/z^2$ power law.}
\label{fig:accel mapping}
\end{figure}
%
It is also interesting to study the effect of spatial scaling in the angular velocity and acceleration of Tornado waves. Let's assume that a Tornado wave is scaled by a factor $s$ so that it's amplitude is now described by $ u(s \cdot r,\varphi)$ (see Eq.~\ref{eq:tornadodef}). Clearly, this is equivalent to a scaling of the radius and  width parameters, of the of the ring-Airy beams that compose the ToW, to $r'_i=r_i/s$, $w'_i=w_i/s$. Using the normalized paraxial wave equation it is straightforward to show that the angular velocity $\upsilon$ and acceleration $\gamma$ will respectively scale as $\upsilon' = s^2 \upsilon$ and $\gamma'=s^4 \gamma$ while the foci will shift to ${f'_i} = {f_i}/s^2$. We can combine all the above to relate the  angular velocity and acceleration to the relative change in the focus position $\tilde f= f'_i/f_i$ so $\upsilon' = \upsilon/\tilde f $ and $\gamma'= \gamma/\tilde f^2$. This practically means that as a ToW is focused tighter ($\tilde f <1$) the angular acceleration is drastically increased following an inverse square power law. This scaling law is expected to hold for all paraxial waves that exhibit angular acceleration and provides a means to fairly compare the angular acceleration $\gamma$ of different types of waves presented in the bibliography. The effect of the focus position on the angular acceleration is clearly shown in Fig.~\ref{fig:accel mapping}. Both cases of ToWs that are generated by ring-Airy beams with foci that partially (rectangles) or completely (stars) overlap are shown. Clearly in the case of foci overlap there is a tenfold increase in the acceleration $\gamma$ compared to the partial foci overlap. As shown Fig.~\ref{fig:accel mapping} ToWs exhibit angular acceleration values that are by orders of magnitude greater than those already reported in the bibliography. This behaviour holds even for the less efficient configuration of partial foci overlap where the $\gamma$ values are by at least  ~5 times higher to the previously reported values \cite{Webster2017,Schulze2015}. {Such high values of angular acceleration are related to the conical nature of the power flow in the foci area. In ring-Airy beams this flow strongly varies along the propagation direction, resulting to angular acceleration. On the other hand, we should note that this comparison is based only in bibliographic data and thus might not fully reflect the range of angular acceleration values that can be achieved by Bessel and Laguerre–Gaussian beams.}

In conclusion, we have shown that by superimposing abruptly auto-focusing ring-Airy beams that carry orbital angular momentum of opposite handedness light spiraling like a tornado can be generated. Using analytical predictions of the abrupt auto-focus position we have tailored the ring-Airy beams so that they abruptly auto-focus at overlapping focal regions. This lead to the generation of a complex wave distribution with intense lobes that twist and shrink in an accelerating fashion along the propagation. Using numerical simulations we have shown that these spiraling waves resemble a tornado, achieving angular acceleration that exceeds $10^4 ~\text{deg/mm}^{2}$. {Although quite complex, Tornado waves can be realized using techniques similar to the ones used for the generation of ring-Airy beams \cite{Papazoglou2011,Panagiotopoulos2013,Efremidis2019}}. Due to their unique features ToWs can be useful to applications ranging from laser trapping to direct laser writing, non-linear wave mixing and harmonics generation, high power THz generation, and filamentation. 

\section*{Funding}
NHQWAVE (MSCA-RISE, 691209); PULSE (Horizon 2020, 824996); EPOCHSE (ERANET, GSRT).

\noindent

\bibliography{Tornado_Beams.bib}

\begin{thebibliography}{10}
\newcommand{\enquote}[1]{``#1''}

\bibitem{Efremidis2010a}
N.~K. Efremidis and D.~N. Christodoulides, {\protect\JournalTitle{Optics
  Letters}} \textbf{35}, 4045 (2010).

\bibitem{Papazoglou2011}
D.~G. Papazoglou, N.~K. Efremidis, D.~N. Christodoulides, and S.~Tzortzakis,
  {\protect\JournalTitle{Optics Letters}} \textbf{36}, 1842 (2011).

\bibitem{Efremidis2019}
N.~K. Efremidis, Z.~Chen, M.~Segev, and D.~N. Christodoulides,
  {\protect\JournalTitle{Optica}} \textbf{6}, 686 (2019).

\bibitem{Penciu2016}
R.-S. Penciu, K.~G. Makris, and N.~K. Efremidis, {\protect\JournalTitle{Optics
  Letters}} \textbf{41}, 1042 (2016).

\bibitem{Panagiotopoulos2013}
P.~Panagiotopoulos, D.~Papazoglou, A.~Couairon, and S.~Tzortzakis,
  {\protect\JournalTitle{Nature Communications}} \textbf{4}, 2622 (2013).

\bibitem{Gauthier2017}
D.~Gauthier, P.~R. Ribi{\v{c}}, G.~Adhikary, A.~Camper, C.~Chappuis, R.~Cucini,
  L.~F. DiMauro, G.~Dovillaire, F.~Frassetto, R.~G{\'{e}}neaux, P.~Miotti,
  L.~Poletto, B.~Ressel, C.~Spezzani, M.~Stupar, T.~Ruchon, and G.~{De Ninno},
  {\protect\JournalTitle{Nature Communications}} \textbf{8}, 14971 (2017).

\bibitem{Yao2011}
A.~M. Yao and M.~J. Padgett, {\protect\JournalTitle{Advances in Optics and
  Photonics}} \textbf{3}, 161 (2011).

\bibitem{Davis2012}
J.~A. Davis, D.~M. Cottrell, and D.~Sand, {\protect\JournalTitle{Optics
  Express}} \textbf{20}, 13302 (2012).

\bibitem{Vasilyeu2009}
R.~Vasilyeu, A.~Dudley, N.~Khilo, and A.~Forbes, {\protect\JournalTitle{Optics
  Express}} \textbf{17}, 23389 (2009).

\bibitem{McGloin2003}
D.~McGloin, V.~Garc{\'{e}}s-Ch{\'{a}}vez, and K.~Dholakia,
  {\protect\JournalTitle{Optics Letters}} \textbf{28}, 657 (2003).

\bibitem{Rop2012}
R.~Rop, A.~Dudley, C.~L{\'{o}}pez-Mariscal, and A.~Forbes,
  {\protect\JournalTitle{Journal of Modern Optics}} \textbf{59}, 259 (2012).

\bibitem{Schulze2015}
C.~Schulze, F.~S. Roux, A.~Dudley, R.~Rop, M.~Duparr{\'{e}}, and A.~Forbes,
  {\protect\JournalTitle{Physical Review A}} \textbf{91}, 043821 (2015).

\bibitem{Webster2017}
J.~Webster, C.~Rosales-Guzm{\'{a}}n, and A.~Forbes,
  {\protect\JournalTitle{Optics Letters}} \textbf{42}, 675 (2017).

\bibitem{Siviloglou2007}
G.~A. Siviloglou, J.~Broky, A.~Dogariu, and D.~N. Christodoulides,
  {\protect\JournalTitle{Physical Review Letters}} \textbf{99}, 213901 (2007).

\bibitem{Siviloglou2007a}
G.~A. Siviloglou and D.~N. Christodoulides, {\protect\JournalTitle{Optics
  Letters}} \textbf{32}, 979 (2007).

\bibitem{Mansour2018}
D.~Mansour and D.~G. Papazoglou, {\protect\JournalTitle{OSA Continuum}}
  \textbf{1}, 104 (2018).

\bibitem{Born1999}
M.~Born and E.~Wolf, \emph{{Principles of Optics: Electromagnetic Theory of
  Propagation, Interference and Diffraction of Light (7th Edition)}} (Cambridge
  University Press, 1999).

\bibitem{Allen}
L.~Allen, M.~W. Beijersbergen, R.~J.~C. Spreeuw, and J.~P. Woerdman,
  {\protect\JournalTitle{Phys. Rev. A}} \textbf{45}, 8185 (1992).

\end{thebibliography}

\end{document}